\documentclass[a4paper,11pt]{article}
\usepackage{jcappub}
\usepackage{natbib}
\usepackage{dcolumn}
\usepackage{graphicx}
\usepackage{hyperref}
\usepackage{color}
\usepackage{bm}
\usepackage{bbold}
\usepackage[usenames,dvipsnames]{xcolor}
\usepackage[normalem]{ulem}
\usepackage{multirow}
\usepackage{ulem}
\usepackage{color}
\usepackage{hyphenat}

\newcommand{\mnras}{Mon.~Not.~R.~Astron.~Soc.}

\newcommand{\prd}{Phys.~Rev. D}
\newcommand{\prc}{Phys.~Rev. C}

\newcommand{\be}{\begin{equation}}
\newcommand{\ee}{\end{equation}}
\newcommand{\bea}{\begin{eqnarray}}
\newcommand{\eea}{\end{eqnarray}}

\title{Constraining models of $f(R)$ gravity with Planck and WiggleZ power spectrum data}
\author[a]{Jason Dossett}
\author[b]{Bin Hu}
\author[a]{David Parkinson}

\affiliation[a]{School of Mathematics and Physics, University of Queensland, \\Brisbane, QLD 4072, Australia}
\affiliation[b]{Institute Lorentz, Leiden University, PO Box 9506, Leiden 2300 RA, The Netherlands}
\emailAdd{j.dossett@uq.edu.au}
\emailAdd{hu@lorentz.leidenuniv.nl}
\emailAdd{d.parkinson@uq.edu.au}

\abstract{
In order to explain cosmic acceleration without invoking ``dark'' physics, we consider $f(R)$ modified gravity models, which replace the standard Einstein-Hilbert action in General Relativity with a higher derivative theory.  We use data from the WiggleZ Dark Energy survey to probe the formation of structure on large scales which can place tight constraints on these models.  We combine the large-scale structure data with measurements of the cosmic microwave background from the Planck surveyor. After parameterizing the modification of the action using the Compton wavelength parameter $B_0$, we constrain this parameter using \texttt{ISiTGR}, assuming an initial non-informative log prior probability distribution of this cross-over scale. We find that the addition of the WiggleZ power spectrum provides the tightest constraints to date on $B_0$ by an order of magnitude, giving  ${\rm log}_{10}(B_0) < -4.07$ at 95\% confidence limit. Finally, we test whether the effect of adding the lensing amplitude $A_{\rm Lens}$ and the sum of the neutrino mass $\sum m_\nu$ is able to reconcile current tensions present in these parameters, but find $f(R)$ gravity an inadequate explanation.}


\begin{document}

\nocite{*}
%\tableofcontents

\maketitle

%% ================================================================ %%
\section{Introduction}

The discovery of the late-time acceleration of the Universe through measurements of Type-Ia supernovae \cite{Riess:1998cb,Perlmutter:1998np,Astier:2006,WoodVasey:2007,Kessler:2009,Conley:2011,Suzuki:2011} and
confirmed by both Cosmic Microwave Background (CMB) \cite{Bennett:2012zja,Ade:2013ktc}
and large-scale structure experiments \cite{Eisenstein:2005,Blake:2011a,Blake:2011b,Anderson:2012sa,Slosar:2013,Busca:2012} is one of the most important discoveries in modern cosmology. 
However, the question of what mysterious force is actually responsible for the acceleration remains an open question. 
Many suggestions have been made, yet at the most fundamental level we are unsure whether the acceleration arises 
from some extra dark fluid component present in the universe or some modification of Einstein's theory of gravity. 
There is a fundamental degeneracy between theories that cannot be broken using only distance data, 
as the dynamics of most general Modified Gravity (MG) theories can easily be replicated by some fluid dark energy 
with an equation of state that varies with scale factor, $w(a)$.  There is therefore a `theory degeneracy' in using only distance data, 
that will require some new form of information to break it.

A large number of different MG theories exist in the literature (for a review, see \cite{Clifton:2012}). 
For example, higher derivative theories such as  $f(R)$ \cite{Carroll:2004,Nojiri:2003} and Galileon \cite{Nicolis:2009} 
models modify the Einstein field equations to be higher than second order. Another type, higher dimension models such as DGP \cite{DGP}, 
change the propagation of the gravity theory by changing the dimensionality of space-time. 
All of these theories have accelerating Friedmann-Lema\^{i}tre-Robertson-Walker (FLRW)-like solutions, which can be tuned to match the distance data without needing 
to introduce a cosmological constant ($\Lambda$). However, by changing the theory of gravity, they affect the motion of particles 
on all scales, beyond the expansion of the homogenous universe. The clustering of matter and growth of large-scale structure 
in the universe is also changed. As such MG theories make very different predictions for the clustering of matter. 
In contrast, most theoretically-motivated dark fluid models (such as 
quintessence) have a very large clustering scale, and so have only a small effect on the 
formation of structure beyond their contributions to the Hubble expansion. In this way, measurements 
of the large scale structure of matter can be used to break this `theory degeneracy' between the MG and dark 
fluid hypotheses, and also distinguish between the different MG theories. 

Cosmological observations have already been used to test and constrain MG theories. 
The cosmic microwave background (CMB) provides a very clean probe of linear structure formation at high redshift, 
but either by nature or by design it is the case that most MG theories make the same predictions at high-redshift as 
the standard concordance model $\Lambda$CDM. 
Even so, measurements of the CMB power spectrum 
and secondary bispectrum 
still have some sensitivity to MG growth of structure through the large-scale integrated Sachs-Wolfe effect and 
weak lensing \cite{Zhang:2005vt,Song:2007,Song:2007da,Ho:2008bz,Marchini:2013lpp,Hu:2013aqa,DiValentino:2012yg,Hu:2012td}. 
At low redshifts, data that have been used included the power spectrum of 
luminous red galaxies \cite{Song:2007,Reyes:2010,Lombriser:2010mp,Yamamoto:2010ie,He:2012wq,Abebe:2013zua}, 
cluster abundances \cite{Schmidt:2009,Ferraro:2010gh,Lombriser:2010mp}, 
Coma cluster \cite{Terukina:2013eqa},
weak lensing \cite{Zhang:2007nk,Hirata:2008cb,Reyes:2010,Daniel:2010ky,Lombriser:2010mp,Tereno:2010dt,Laszlo:2011sv,Simpson:2013},
redshift-space distortions \cite{Simpson:2013,Jennings:2012pt,Raccanelli:2012gt}, 21 cm line \cite{Hall:2012wd,Wang:2010ug}
and matter bispectrum \cite{GilMarin:2011xq,Bartolo:2013ws}.

In this paper, we choose to test the $f(R)$ gravity theory, as it is one of the simplest theories available, and since the function of the Ricci scalar $R$ can be chosen, it can be ``tuned" to reproduce any background expansion history needed \cite{Song:2006ej,Pogosian:2007sw,Hu:2013twa}. We make predictions of the growth of structure under this theory, and compare them to observation to constrain the associated parameters. To do this, we use measurements of the galaxy power spectrum made by the WiggleZ Dark Energy Survey \cite{Drinkwater:2010} and 
combine it with recent measurements of the CMB power spectrum from the Planck surveyor \cite{Ade:2013ktc}. The galaxy power spectrum from the WiggleZ Dark Energy Survey has advantages over some other large-scale survey data, as the effect of non-linearites (redshift-space distortions and non-linear structure formation)  seem to be small \cite{Parkinson:2012}.

The outline of our paper is as follows. 
In section \ref{sec:Theory} we describe the theoretical basis of the $f(R)$ theory we are considering and the predictions that it makes.  
In section \ref{sec:datasets} we describe the different data sets  we used in the analysis. 
In section \ref{sec:analysis} we give the results from our likelihood analysis, and discuss them.
Finally, we conclude in section \ref{sec:conclusions}. 

%% ------------------------------------------------------------------ %%
\section{Theory}\label{sec:Theory}

In this section, we will describe the theoretical framework in testing gravity by using cosmological data. 
First, we will introduce the generic formalism in the subsection \ref{sec:Theory1}. Then, in 
subsection \ref{Sec:theory2} we will study one of the 
explicit parameterizations of $f(R)$ gravity under the quasi-static (QS) approximation. 
\subsection{The growth of perturbations}\label{sec:Theory1}
Let us consider perturbations of the flat FLRW metric in the conformal Newtonian Gauge.  In this gauge the metric is written
\be
ds^2=a(\tau)^2[-(1+2\psi)d\tau^2+(1-2\phi)dx^idx_i],
\label{eq:FLRWNewt}
\ee
where $\tau$ is conformal time, $a(\tau)$ is the scale factor normalized to one today, $\psi$ and $\phi$ are the potentials describing the scalar modes of the metric perturbations,  and the $x_i$'s are the comoving coordinates.

Applying Einstein's field equations to this metric, one can obtain the Poisson and anisotropy equations. The first comes from combining the time-time and time-space equations and the second from the traceless space-space equation.  In the context of testing general relativity, these equations are modified to include terms which will mimic the effects of a modified gravity model on the growth in the linear regime. The modified Poisson and anisotropy equations as written in the formalism of \cite{Bean:2010zq} are given, respectively, by 
\bea
k^2\phi  &=& -4\pi G a^2\sum_i \rho_i \Delta_i \,  Q(k,a)
\label{eq:PoissonMod}\\
k^2(\psi-R(k,a)\,\phi) &=& -12 \pi G  a^2\sum_i \rho_i (1+w_i) \sigma_i \, Q(k,a),
\label{eq:Mod2ndEin}
\eea 
where $\sigma_i$ is the shear stress, and $\rho_i$ is the density (with $_i$ denoting a particular matter species).  
The modified growth functions, $Q(k,a)$ and $R(k,a)$, can have both time (described in terms of scale factor, $a$) and scale dependence. 
A modification to the Poisson equation is quantified by the parameter $Q(k,a)$, while $R(k,a)$ quantifies the so-called gravitational slip 
(a term coined by \cite{CCM07} to refer to the ratio between the two metric potentials) as at late times, assuming negligible anisotropic stress
from normal matter components, $\psi= R\phi$.  
It is worth noting that eq.\ \eqref{eq:PoissonMod}, above, cannot truly be called the Poisson equation, as it often is.  
It relates the space-like potential, $\phi$, (the one only affecting relativistic particles) to the overdensity, 
while the Poisson equation should relate the  overdensity to a potential which influences the dynamics of all particles. 
The potential that does this is the time-like, Newtonian potential, $\psi$.

A similar modified growth formalism had been introduced by \cite{Bertschinger:2008zb} prior to the formalism described above.  This formalism however was introduced under the assumption of negligible anisotropic stress and therefore could only be applied at late times.  Under these assumptions the modified Poisson and anisotropy equations are recast to include two observation-related variables, $\mu(a,k)$, which defines a time and scale-dependent Newton's constant through the product $\mu(a,k)G$, and the gravitational slip $\gamma(a,k)$.  These equations are written
\bea
\label{BZ}
k^2\psi &=& -4\pi G\, \mu(a,k)a^2\rho\Delta\;,\\
\label{BZ1.2}
\frac{\phi}{\psi}&=&\gamma(a,k)\;.
\eea
As pointed out in \cite{Dossett:2011fm}, functions $Q$ and $R$ are simply related to $\mu$ and $\gamma$ in the limit of negligible matter anisotropic stress, $\sigma$ via the following relations:
\be
\label{BZ_VS_TB}
Q=\mu\gamma\;,\quad R=\gamma^{-1}\;.
\ee
In this paper, we primarily use the extended formalism eq.\ (\ref{eq:PoissonMod}) and (\ref{eq:Mod2ndEin}). 
 
Above, $\Delta_i = \delta_i+3\mathcal{H}(1+w_i)\theta_i/k^2$ is the comoving overdensity, with $\delta_i$ the overdensity, $\theta_i$ the divergence of the peculiar velocity, and $\mathcal{H}$ the Hubble parameter in conformal time.  Enforcing conservation of energy momentum on a perturbed fluid gives the evolution of these quantities as \cite{Ma:2005}:
\be
\delta'  =  -(1+w)(\theta-3\phi')+3\mathcal{H}(w-\frac{\delta P}{\delta\rho})\delta
\label{eq:deltaevo}
\ee
\vspace{-\baselineskip}
\be
\theta'= -\mathcal{H}(1-3w)\theta-\frac{w'}{1+w}\theta +\frac{\delta P/\delta\rho}{1+w}k^2\delta+k^2(\psi-\sigma).
\label{eq:thataevo}
\ee
where $w$ is the equation of state which relates the pressure of a fluid, $P$ to its density, $\rho$ via the usual relation $P = w \rho$ and primes denote derivatives with respect to conformal time, $\tau$.

On linear scales, the growth of cosmological structures can be described almost entirely by the growth of 
the overdensity for cold dark matter (CDM), $\delta_m$.  Taking into account that CDM is pressureless and shear free, 
we can use the above two equations and (after switching to proper time) obtain the usual equation for the growth:
\be
\ddot{\delta}_m+2H\dot{\delta}_m+\frac{k^2\psi}{a^2}=0.
\label{eq:delgrow}
\ee
where dots denote derivatives with respect to proper time, $t$.

In the GR case, subbing in for $k^2 \psi$ using eqs.\ \eqref{eq:PoissonMod} and \eqref{eq:Mod2ndEin} gives the even more familiar equation
\be
\ddot{\delta}_m+2H\dot{\delta}_m-4\pi G \rho_m \delta_m=0,
\label{eq:delgrowGR}
\ee
which is independent of scale, $k$.  This scale independence does not necessarily hold for modified gravity models. 
Such is the case for the $f(R)$ modified gravity models that we discuss below.

%% ------------------------------------------------------------------ %%
\subsection{$f(R)$ gravity theories\label{Sec:theory2}}
Due to the simplicity of its Lagrangian, $f(R)$ gravity obtained a lot of attention,
(see the recent review \cite{DeFelice:2010aj} and references therein) especially as an 
illustration of the chameleon mechanism. Besides the simplicity of the structure of this theory, 
there exist some other reasons for the interest it attracted. 
First of all, the form of the function $f(R)$ can be engineered to exactly mimic {\it any} 
background history via a one-parameter family of solutions~\cite{Song:2006ej,Pogosian:2007sw,Hu:2013twa}. 
Second, $f(R)$ gravity can provide a slightly better fit to the CMB data than flat $\Lambda$CDM, 
which can be attributed to the lowering of the temperature anisotropy power spectrum at the low-multipole
regime~\cite{Lombriser:2010mp}. 
Its Lagrangian in the Jordan frame could be written as 
\begin{equation}
S = \int d^4x \sqrt{-g} \left[ \frac{R+f(R)}{16\pi G} + \mathcal{L}_m \right] \,,
\end{equation}
where $\mathcal{L}_m$ is the minimally coupled matter sector.

Because of the higher order derivative nature of $f(R)$ gravity, there exists a scalar degree of freedom, named the scalaron
$f_R\equiv{\rm d} f/{\rm d} R$ with mass
\be\label{fR_mass}
m^2_{f_R}\equiv\frac{\partial^2V_{\rm eff}}{\partial f^2_{R}}=\frac{1}{3}\left(\frac{1+f_R}{f_{RR}}-R\right)\;.\ee
The corresponding Compton wavelength reads
\be\label{comp_wav1}
\lambda_{f_R}\equiv m^{-1}_{f_R}\;.\ee
Usually, it is convenient to use the dimensionless Compton wavelength in Hubble units
\be\label{comp_wav2}
B\equiv\frac{f_{RR}}{1+f_R}R'\frac{H}{H'}\;,\ee
with $f_{RR}={\rm d}^2 f/{\rm d} R^2$ and ${~}'={\rm d} /{\rm d} \ln a$. 
In the GR limit, the scalar field disappears due to the infinite mass, i.e. 
zero Compton wavelength ($B(a)=0$). 
Thanks to this extra scalar degree of freedom as well as the background symmetries (homogeneity and isotropy), 
on the background level the function form of $f(R)$ could be engineered to 
mimic the given expansion history \cite{Song:2006ej,Pogosian:2007sw,Hu:2013twa}. 
Hence, we could not distinguish dark energy models via the background kinematic tests alone. 
Fortunately, at the perturbation level breaking of homogeneity and isotropy will also break 
this theoretical degeneracy, i.e.\ we could recognize different dark energy models via their growth structure dynamics. 

Armed with these theoretical observations, let us now turn to the $Q(k,a)$ and $R(k,a)$ parameterization in $f(R)$ gravity.  In general, a reasonable parameterization is usually based on some approximations. Here, it is the quasi-static (QS) assumption, which means that we only keep the spatial derivative terms in equations while neglecting the temporal derivatives. 
For a small deviation from GR ($B(a)\ll 1$), it has been proven that the 
QS description for $f(R)$ gravity is satisfactory for several ongoing projects \cite{Hojjati:2012rf}. 
This is because in $f(R)$ gravity, aside from the horizon scale, in the linear regime there exists a new scale --- the Compton scale --- which characterizes the deviation from GR. Above this scale, its deviation is tiny; while below 
it, the modified dynamics could be obvious. When the Compton scale sits deeply inside the horizon, i.e.\ small $B(a)$ case, 
the information of modified gravity is mainly characterized by the sub-horizon dynamics, the temporal evolution of which, is
insignificant. Hence, for this case the QS approximation holds. However, when the Compton scale is comparable with horizon 
scale ($B(a)\sim 1$), the temporal evolution of gravitational potentials are no longer negligible and the QS assumption breaks down. 
From a practical observational point of view, the current cosmological probes already rule out the large Compton wavelength case. 
For example, the recently released Planck CMB temperature and lensing power spectra data gives the present Compton wavelength bound
$B_0<0.1$ at $95\%$C.L. \cite{Hu:2013aqa} and the joint analysis of several LSS tracers combined with WMAP data gives a even more stringent constraint 
$B_0<1.1\times10^{-3}$ at $95\%$C.L. \cite{Lombriser:2010mp}.  
Given the above theoretical explanations and current data analysis status, we will adopt this QS approximation in our following study. 

For $f(R)$ modified gravity models, by assuming quasi-staticity and $\Lambda$CDM background,
the functional form of the modified growth parameters, $Q(k,a)$ and $R(k,a)$, has been given by 
\cite{Bertschinger:2008zb, Tsujikawa:2007, Giannantonio:2009gi} and can be written as:
\be
Q(k,a) = \frac{1}{1-1.4\times 10^{-8}\lambda_1^2a^3}\frac{1+\frac{2}{3}\lambda_1^2k^2a^s}{1+\lambda_1^2k^2a^s},\label{eq:QFR}
\ee
\be
R(k,a) = \frac{1+\frac{4}{3}\lambda_1^2k^2a^s}{1+\frac{2}{3}\lambda_1^2k^2a^s},\label{eq:RFR}
\ee
where the empirical prefactor in eq.\ (\ref{eq:QFR}) corresponds to corrections to more accurately model the ISW contributions in $f(R)$ gravity \cite{Giannantonio:2009gi}. 
And $\lambda_1$ is nothing but the present Compton wavelength $\lambda_1^2 = B_0c^2/(2H_0^2)$. 
As demonstrated in \cite{Hu:2013twa}, this parameterization assumes a power-law growth of the 
Compton wavelength
\begin{align}
\frac{f_{\rm RR}}{1+f_{\rm R}}= \frac{B_0}{6H_0^2}a^{s+2}.
\label{Eq:BZansatz}
\end{align}
We should emphasize that a constant value for $s$ will not in general be capable of reproducing the full $\Lambda$CDM expansion history.
However, it works as a good approximation for each epoch alone~\cite{Thomas:2011pj}, 
as can be inferred from eq.\~(\ref{Eq:BZansatz}). 
Indeed a reasonable value of $s$ is given by $s\approx 5$ during radiation domination, $s\geq4$  
during matter domination and $s<4$ during the late time phase of accelerated expansion.
For small values of $B_0$, it is customary to fix  $s=4$ as discussed in~\cite{Hojjati:2012rf}. 
In this paper we adopt this choice and then play with a single parameter $B_0$. 

Combining the above eqs.\ \eqref{eq:QFR} and \eqref{eq:RFR} with eqs.\ \eqref{eq:PoissonMod} and \eqref{eq:Mod2ndEin} and subbing into eq.\ \eqref{eq:delgrow} gives the following equation for the growth of matter perturbations in linear regime in $f(R)$ gravity:
\be
\ddot{\delta}_m+2H\dot{\delta}_m- \frac{4\pi G}{1-1.4\times 10^{-8}\lambda_1^2a^3}\frac{1+\frac{4}{3}\lambda_1^2k^2a^4}{1+\lambda_1^2k^2a^4}\rho_m \delta_m=0.
\label{eq:delgrowFR}
\ee
In contrast to eq.\ \eqref{eq:delgrowGR} which is scale independent, growth in $f(R)$ modified gravity models is dependent upon scale as seen above. More useful is the fact that the growth of matter perturbations are dependent upon the the value of $\lambda_1^2$ or equivalently $B_0$.  A larger $B_0$ will boost growth, albeit in a scale-dependent way. As such cosmological observations which probe the growth of matter perturbations, such as the matter power spectrum (MPK) should be useful in placing constraints on the $f(R)$ parameter $B_0$.

%% ================================================================ %%
\section{Datasets}
\label{sec:datasets}

We use the measurements of CMB temperature anisotropy\footnote{\url{http://pla.esac.esa.int/pla/aio/planckProducts.html}} \cite{Ade:2013zuv}
from the first data release of the Planck surveyor. 
Its temperature power-spectrum likelihood is divided into low-$l$ ($l<50$) and high-$l$ ($l\geq 50$) parts. This is because the central limit theorem ensures that the distribution of CMB angular power spectrum $C_l$ in the high-$l$ regime can be well approximated by a Gaussian statistics. However, for the low-$l$ part the $C_l$ distribution is non-Gaussian. 
For these reasons the Planck team adopts two different methodologies to build the likelihood. In detail,
for the low-$l$ part, the likelihood exploits all Planck frequency channels
from $30$ to $353$ GHz, separating the cosmological CMB signal from diffuse Galactic foregrounds through a physically motivated Bayesian component separation technique. 
For the high-$l$ part, a correlated Gaussian likelihood approximation is employed. This is based on a fine-grained set of angular cross-spectra derived from multiple detector power-spectrum combinations between the $100$, $143$, and $217$ GHz 
frequency channels, marginalizing over power-spectrum foreground templates. In order to break the well-known parameter degeneracy between the reionization optical depth $\tau$ and the scalar index $n_s$, the low-$l$ WMAP polarization likelihood (WP) is used\cite{Ade:2013zuv}. 
Finally, the unresolved foregrounds are marginalized over, assuming wide priors on the relevant nuisance parameters as described in \cite{Planck:2013kta}.

In order to take advantage of the constraining power of data from large-scale structure, we use measurements of the galaxy power spectrum as made by the WiggleZ Dark Energy Survey\footnote{\url{http://smp.uq.edu.au/wigglez-data}}. As described in \cite{Parkinson:2012}, we use the power spectrum measured from spectroscopic redshifts of 170,352 blue emission line galaxies over a volume of ~1 Gpc$^3$ \cite{Drinkwater:2010}. The covariance matrices as given in \cite{Parkinson:2012} are computed using the method described by \cite{Blake:2010}.  The best model proposed for non-linear corrections to the matter power spectrum was one that was calibrated against simulations (model G in \cite{Parkinson:2012}), and so this model may not be appropriate for this situation. Instead, it has already been demonstrated that linear theory predictions are as good a fit to the data as the calibrated model to $k \sim 0.2 h/{\rm Mpc}$  \cite{Parkinson:2012,Riemer-Sorensen:2013}. Furthermore, recent work \cite{Zhao:2013} has shown that the linear power spectrum produced by codes using the modified growth formalism detailed above reproduce quite accurately the power-spectrum obtained using N-Body simulations for $f(R)$ gravity models out to $k \sim 0.2 h/{\rm Mpc}$.  The linear predictions, in fact, are more accurate than those made using the non-linear matter power spectrum fitting formula module, Halofit. For these reasons we restrict ourselves to scales less that $k_{\rm max} = 0.2 h/{\rm Mpc}$ and use the linear theory prediction only.  We do investigate the effect of a different cut-off scale of $k_{\rm max} = 0.1 h/{\rm Mpc}$ on our results, to test for possible non-linear systematic error. We also marginalise over a linear galaxy bias for each of the four redshift bins, as in \cite{Parkinson:2012}.

Finally, in order to break other parameter degeneracies relating to late-time observables such as $\Omega_m$ we also use baryon acoustic oscillation (BAO) measurements from the 6dF Galaxy Survey measurement at $z= 0.1$ \cite{Beutler:2011hx}, the re-analyzed SDSS DR7 \cite{Padmanabhan:2012hf,Percival:2009xn} at effective redshift $z_{\rm eff}=0.35$, and the BOSS DR9 \cite{Anderson:2012sa} surveys at $z_{\rm eff}=0.2$ and $z_{\rm eff}=0.35$.

%% ================================================================ %%
\section{Results \& Discussion}
\label{sec:analysis}
We used a Markov-Chain Monte Carlo method to obtain posterior constraints on the parameters, 
using two separate modifications of the cosmological analysis code \texttt{CosmoMC} \cite{Lewis:1999bs,Lewis:2002ah}, namely \texttt{ISiTGR}\footnote{\url{http://isit.gr}} \cite{Dossett:2011tn,Dossett:2012kd} and \texttt{MGCAMB} \cite{Zhao:2008bn,Hojjati:2011ix}. Since the constraints from the two codes were quite consistent, in what follows we primarily show the results from \texttt{ISiTGR}. In table \ref{tab:params} we list the priors on the various parameters used in our analysis.  

\setlength\tabcolsep{1pt}
\begin{table*}[htb!]
\footnotesize
\centering
\begin{tabular}{|l|c|c|c|}
\hline
{\bf Cosmological Parameter} & symbol & \multicolumn{2}{c|}{\underline{~~~~prior~~~~}}  \\ 
 & & min & max \\ \hline
 {\bf \textit {$\Lambda$CDM parameters}} & & & \\
Physical baryon density & 100$\Omega_{\rm b} h^2$  & 0.5 & 10. \\
Physical CDM density & $\Omega_{\rm CDM} h^2$ &0.001 & 0.99 \\
Angular size of the sound horizon at decoupling  & 100$\theta$& 0.5 & 10 \\
Optical depth of reionization & $\tau$ &0.01 & 0.8 \\
Scalar spectral index & $n_s$ & 0.9 & 1.1 \\
Amplitude of the scalar perturbations & $\log(A_s)$ & 2.7 & 4 \\
\hline
{\bf \textit{Parameters to extend the model}} &  & &\\
Compton wavelength of the $f(R)$ theory & $\log(B_0)$ & -10 & 1 \\
CMB lensing parameter & $A_{\rm Lens}$ & 0 & 10  \\
Combined mass of the neutrino species (eV) & $\sum m_{\nu}$ & 0 & 5 \\ 
\hline
{\bf \textit{Derived parameters}} &   & & \\
Matter density & $\Omega_m$ & -- & -- \\
Hubble parameter (kms$^{-1}$Mpc$^{-1}$) & $H_0$  & -- & -- \\
Matter power dispersion at $8 h^{-1}$Mpc & $\sigma_8$ &-- & -- \\ 
\hline
\end{tabular}
\caption{\label{tab:params}Cosmological parameters and prior ranges.}

\end{table*} 

The logarithmic prior on $B_0$ is a non-informative prior, which gives equal weighting to the largest values allowed for the Compton wavelength (${\rm log}_{10}(B_0) = 1$ is equivalent to the gigaparsec scale) and the smallest values (${\rm log}_{10}(B_0) = -10 \sim$ tens of kiloparsecs). In contrast a uniform prior on $B_0$ would give additional probability weight to the largest scales, and would possibly bias the result. We do not expect to be able to detect the Compton wavelength on the kiloparsec scale, but assume this as a reasonable lower limit given we observe no modified gravity signal on galactic scales.

Throughout this work we will use three different combinations of data sets.  The first combination uses Planck + WP + BAO and is denoted PLC (standing for \textit{Plank likelihood code}) hereafter.  Next, in addition to the PLC data we add data from the WiggleZ galaxy power spectrum with data points out to $k_{\rm max}=0.1h/{\rm Mpc}$ We denote this data set as PLC + WiggleZ$_{0.1}$.  Finally, we include data from the WiggleZ galaxy power spectrum out to $k_{\rm max}=0.2h/{\rm Mpc}$ and denote this combination of data sets PLC + WiggleZ$_{0.2}$.

We analyze constraints on the parameters for three different models as well: 
\begin{itemize}
\item{ \textbf{Model I} - $f(R)$ gravity only, adding the parameter $\rm log_{10}(B_0)$;}
\item{ \textbf{Model II} - $f(R)\, +\, A_{\rm  Lens}$, where we vary the additional CMB lensing amplitude parameter $A_{\rm Lens}$; and}
\item{ \textbf{Model III} - $f(R)\, +\, A_{\rm  Lens}\, +\, \sum m_\nu$, where we also vary the sum of the mass of active neutrino species.}
\end{itemize}  

Note that we assume three massive, degenerate active neutrinos, and in Models I \& II where the mass remains fixed we assume a value of $\sum m_\nu = 0.06~{\rm eV}$. 

%% ------------------------------------------------------------------ %%
\subsection{Model I - $f(R)$ gravity}

% *-*-*-*-*-*-*-*-*-*-*-*-*-*-*-*

% *-*-*-*-*-*-*-*-*-*-*-*-*-*-*-*
\begin{figure*}[ht!]
\centering
\includegraphics[scale=0.35]{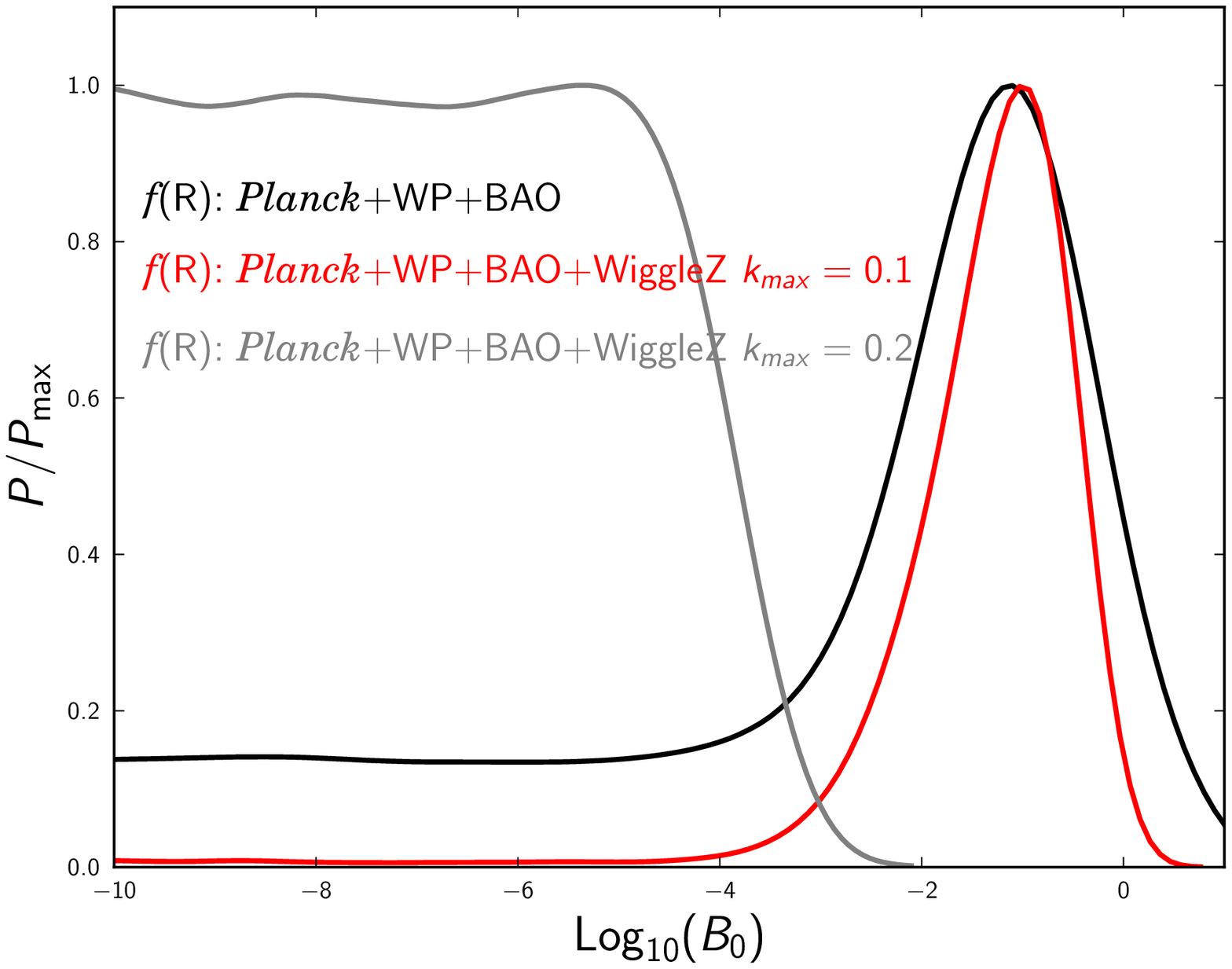}
\includegraphics[scale=0.35]{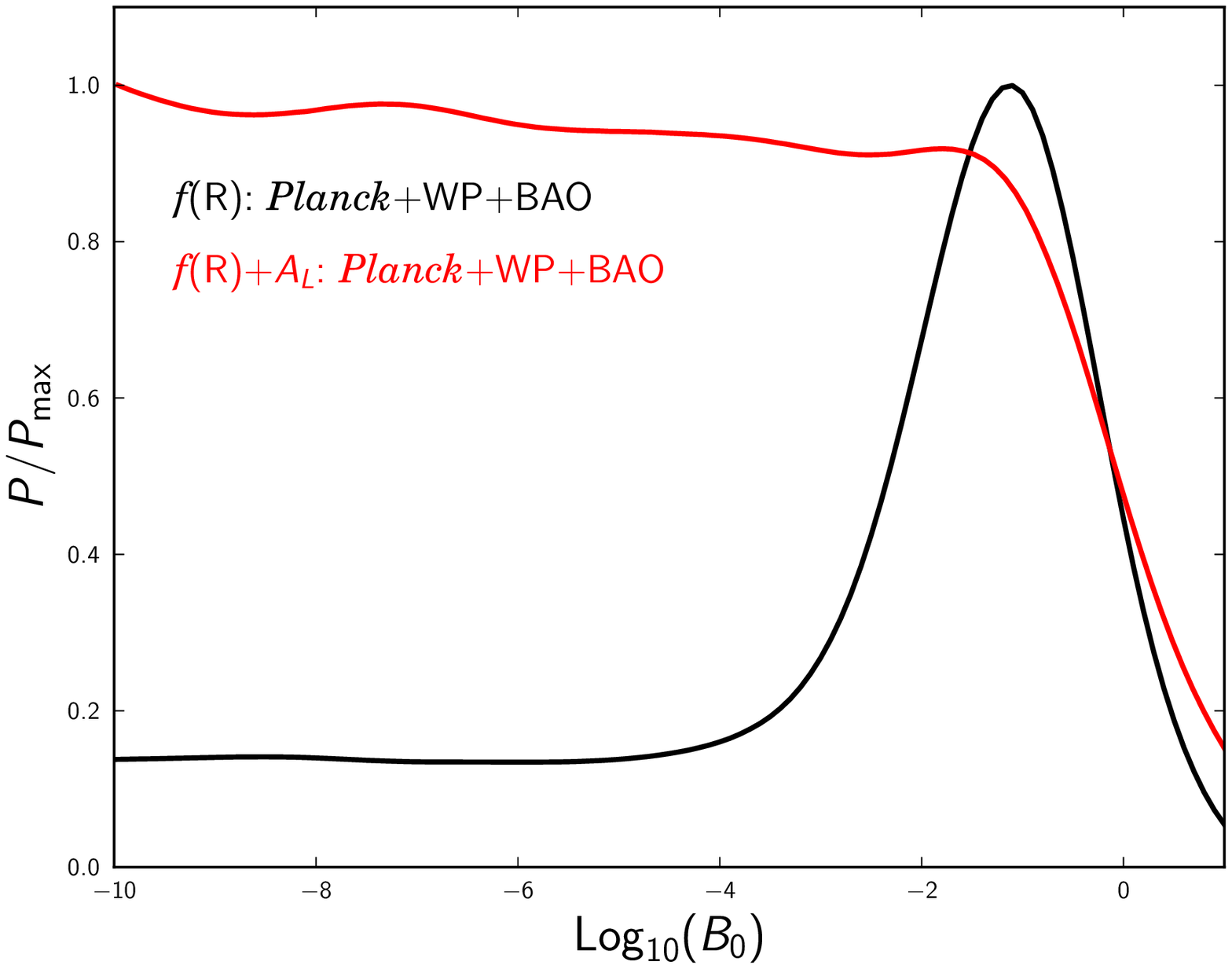}
\caption{\label{f:f(R)A1D}  
{\it Left panel:} 1D posterior distribution of ${\rm log}_{10}(B_0)$ for our three different combinations of data sets.  The PLC + WiggleZ$_{0.1}$ data set yeilds a $3\sigma$ detection of a non-zero $B_0$.  Extending the data points of the WiggleZ data set up to $k_{\rm max} = 0.2h/{\rm Mpc}$, however, removes this detection and places a very stringent constraint on $B_0$.  {\it Right panel:}  The 1D posterior distribution of ${\rm log}_{10}(B_0)$ for Model I and II for the PLC data set. The high-probability peak at larger values of $B_0$ for the PLC data set vanishes the addition of the $A_{\rm Lens}$ parameter, as discussed in section \ref{sec:fr_alens}.}

\end{figure*}
% *-*-*-*-*-*-*-*-*-*-*-*-*-*-*-*

% *-*-*-*-*-*-*-*-*-*-*-*-*-*-*-*
\begin{figure}
\centering
\begin{tabular}{c}
{\includegraphics[width=8cm]{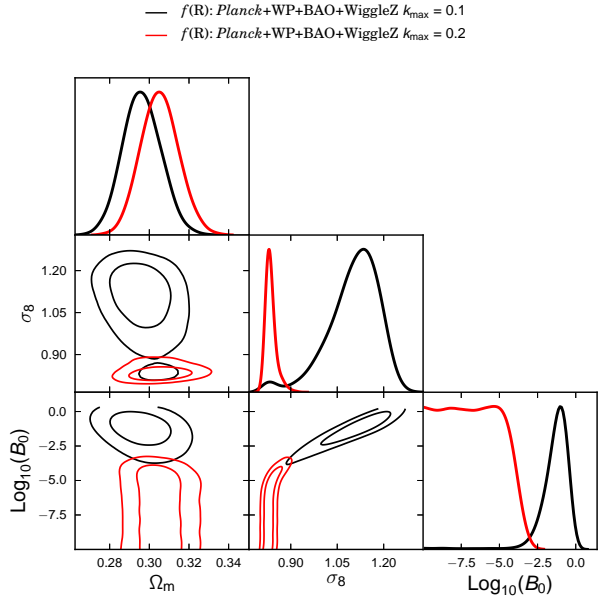}} \\ 
\end{tabular}
\caption{\label{f:2d_fr_wigglez_planck} Plot showing joint constraints on cosmological parameters from combining 
Planck CMB and WiggleZ galaxy power spectra data. 
The WiggleZ data is fit up to $k_{\rm max}=0.1 h/{\rm Mpc}$ (black) and up to  
$k_{\rm max}=0.2 h/{\rm Mpc}$ (red).}
\end{figure}
% *-*-*-*-*-*-*-*-*-*-*-*-*-*-*-*

We first consider the cosmological constraints on $f(R)$ gravity alone, 
with no additional, beyond the standard model, parameters.  Our constraints on the $f(R)$ parameter ${\rm log}_{10}B_0$ are given in figure \ref{f:f(R)A1D}. We can quickly see that the that the addition of galaxy power spectrum form WiggleZ can improve the $95\%$ upper bound of $B_0$ by over three orders of magnitude
\begin{eqnarray}
 {\rm log}_{10}(B_0) &<& -0.44\;, (95\%,\, \mbox{ Model I:}\;{\rm PLC}),\label{eq:f(R)PLC}\\
 {\rm log}_{10}(B_0) &<& -4.07\;, (95\%,\, \mbox{ Model I:}\;{\rm PLC}+{\rm WiggleZ_{0.2}}).\label{eq:f(R)WIGZk0p2}
\end{eqnarray}
This represents one of the tightest constraints $B_0$ and thus $f(R)$ models to date as previous constraints from CMB data sets alone gave $B_0<0.1$ at $95\%$C.L. \cite{Hu:2013aqa}, and joint analysis of several large-scale structure tracers combined with CMB data from WMAP had given constraints of $B_0<1.1\times10^{-3}$ at $95\%$C.L \cite{Lombriser:2010mp}.

The most noticeable results, as seen in the left panel of figure \ref{f:f(R)A1D}, are the two high probability peaks for non-zero values for $B_0$ for the PLC and PLC + WiggleZ$_{0.1}$ data set combinations.  For the PLC data set this peak does not represent a significant detection of a non-zero $B_0$ and is removed by adding $A_{\rm Lens}$ to the parameter analysis as discussed more in the next section. However, for PLC + WiggleZ$_{0.1}$ this represents a more than $2\sigma$ detection of a non-zero $B_0$ as
\begin{equation}
 {\rm log}_{10}(B_0) = -1.42^{+1.55}_{-1.66}~(95\%,\, \mbox{ Model I:}\;{\rm PLC}+{\rm WiggleZ_{0.1}}).\label{eq:f(R)WIGZk0p1}
\end{equation}
We argue that this is due to the fact that for $k_{\rm max} = 0.1h/{\rm Mpc}$  the power spectra data has a slight preference for a lower value $\Omega_m$, and hence a higher value of $\sigma_8$ and $B_0$ are needed to accommodate a fit to the CMB.  This is illustrated well in figure \ref{f:2d_fr_wigglez_planck} where we plot the two-dimensional joint posterior parameter distributions between the matter density, $\Omega_m$, the amplitude of clustering, $\sigma_8$, and the $f(R)$ parameter, ${\rm log}_{10}(B_0)$. This difference in best fit between WiggleZ$_{0.1}$ and WiggleZ$_{0.2}$ comes from the scale-dependent growth in $f(R)$. In the $f(R)$ model power on small scales is boosted, but when this combined with our free linear bias parameter, it results in a better fit at larger scales by predicting a slightly lower power for the small k-values. However, the scale-dependent growth of $f(R)$ provides a worse fit for the smaller scale data, and so when we include the data out to $k_{\rm max}=0.2 h/{\rm Mpc}$, the data disfavors larger values of the Compton wavelength.

% *-*-*-*-*-*-*-*-*-*-*-*-*-*-*-*
\begin{figure}[b]
\centering
\begin{tabular}{c}
{\includegraphics[width=8cm]{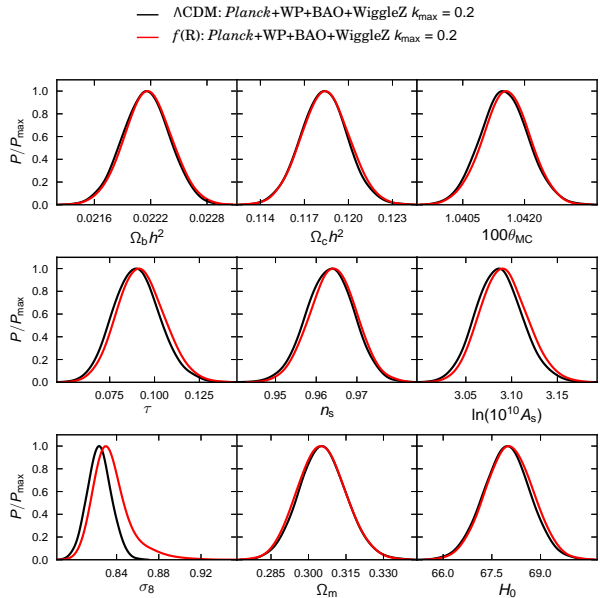}} \\ 
\end{tabular}
\caption{\label{f:WIGZdata1D} Plot showing 1D posterior probability distributions for 
the standard $\Lambda$CDM parameters and derived parameters, as described in  table \ref{tab:params}, 
comparing the cases where the standard Einstein gravity is assumed (black) and the $f(R)$ gravity is used (red). 
We see almost no difference between the distributions, except a low-probability tail for large values of the clustering 
amplitude $\sigma_8$ in the $f(R)$ case.}
\end{figure}
% *-*-*-*-*-*-*-*-*-*-*-*-*-*-*-*

Finally, in figure \ref{f:WIGZdata1D}, we plot the 1D probability distributions for the core cosmological parameters using the WiggleZ power spectrum out to $k_{\rm max}=0.2$. With the exception of the low-probability tail for $\sigma_8$, the best fit values for the standard cosmological parameters do not change much compared to the $\Lambda$CDM case.  This is because the constraints on most of the core cosmological parameters come from the CMB, and are very well-constrained by Planck. We found this to be generally the case throughout our analysis. For the purposes of completeness, the means and standard deviations of all the cosmological parameters used in our fits are given in table \ref{Tab:foR} in appendix \ref{App:paramconstraints}.

% *-*-*-*-*-*-*-*-*-*-*-*-*-*-*-*
\begin{figure*}[h!]
\centering
\includegraphics[scale=0.4]{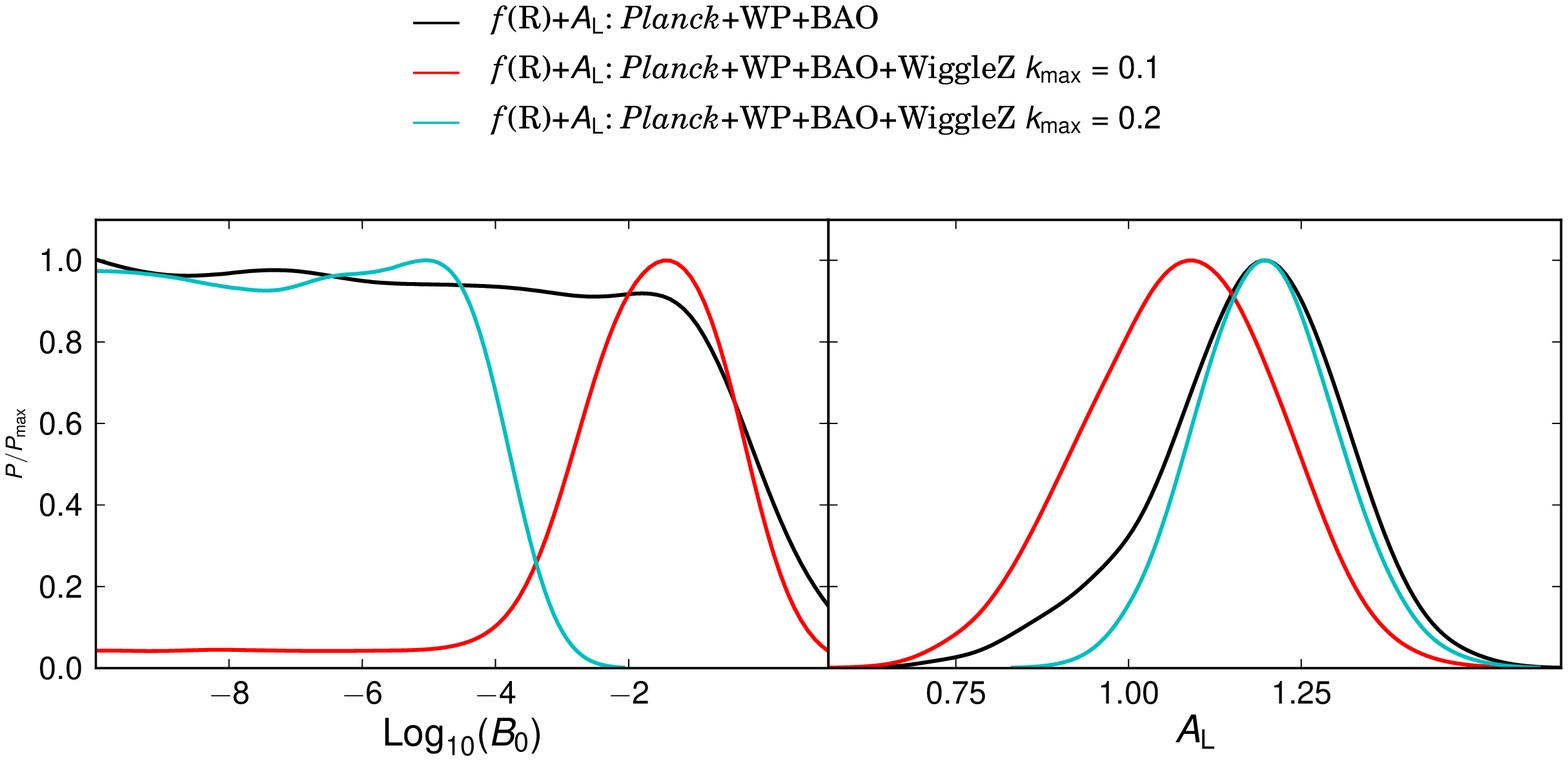}
\caption{\label{f:1Df(R)ALB0AL}1D posterior distributions from Model II for our three different combinations of data sets. {\it Left panel:} Distributions for ${\rm log}_{10}(B_0)$. The non-zero detection of $B_0$ still persists for the ${\rm PLC}+{\rm WiggleZ_{0.1}}$ data set, while the PLC data set no displays a high probability peak corresponding to non-zero values of $B_0$.  The PLC+WiggleZ$_0.2$ data set still provides a very stringent constraint on $B_0$.
{\it Right panel:} 1D posterior distributions for $A_{\rm Lens}$.  Here the PLC + WiggleZ$_{0.1}$ (red curve) shows a preference for $A_{\rm Lens}$ very close to its expected value of one.  This is not the case for the other two data sets which prefer values for $A_{Lens}$ which deviate from unity.}
\end{figure*}
% *-*-*-*-*-*-*-*-*-*-*-*-*-*-*-*

%% ------------------------------------------------------------------ %%
\subsection{Model II - $f(R)$ gravity + $A_{\rm Lens}$}\label{sec:fr_alens}

We now consider models where in addition to the $f(R)$ model parameter ${\rm log}_{10}(B_0)$ we vary the parameter $A_{\rm Lens}$, the normalized the CMB lensing amplitude.  This parameter has garnered a lot of attention since the Planck 2013 data release where a $2\sigma$ deviation from unity $A_{\rm Lens}=1.23\pm 0.11$ was found. This is significant because a deviation in the value of this parameter from unity indicates a tension with $\Lambda$CDM cosmology.  On one hand this tension could be due to some unknown systematics in the Plank power spectrum data, and thus be resolved by better understanding the experiment and resolving said systematics.  
In theory, however, this tension could be reconciled by considering models beyond standard six parameter $\Lambda$CDM.  Some work has already been done to this end (see, for example, \cite{Marchini:2013oya,Hu:2014qma}). It is for this reason we explore values of this parameter in $f(R)$ gravity. 

We provide the means and standard deviations of the parameters for this model in table \ref{Tab:foR+AL} in appendix \ref{App:paramconstraints} and focus more on the plotted results for this model which are given in figure \ref{f:1Df(R)ALB0AL} for our three data set combinations.  From the left hand side of this figure we can quickly see that the addition of the parameter $A_{\rm Lens}$ removes the high probability peak corresponding to larger values of $B_0$ that was seen when using only the PLC data.  At the same time, it weakens the constrains so much that there is no upper bound on $B_0$ for the priors we have selected, which is consistent with the results in \cite{Hu:2013aqa}. Also, while the constraints are obviously wider than in Model I, the detection of a non-zero $B_0$ is still present when using $\mbox{PLC + WiggleZ}_{0.1}$.  The data set combination $\mbox{PLC + WiggleZ}_{0.2}$, though, is still able to provide a very stringent constrain on $B_0$ and thus $f(R)$ gravity with a mildly wider upper 
bound due to the inclusion of $A_{\rm Lens}$ in the analysis of
\begin{equation}
 {\rm log}_{10}(B_0) < -4.02, (95\%,\, \mbox{Model II:}\;{\rm PLC}+{\rm WiggleZ}_{0.2}).\label{eq:f(R)AWIGZk0p2}
\end{equation}

% *-*-*-*-*-*-*-*-*-*-*-*-*-*-*-*
\begin{figure*}[tmb]
\centering
\includegraphics[scale=0.5]{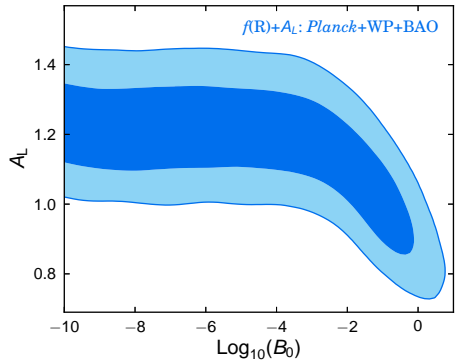}
\caption{We plot the 2D contour posterior distribution for ${\rm log}_{10}B_0$ and $A_{\rm Lens}$ when using the PLC data set.  We can see a large degeneracy between these parameters at large values of $B_0$.}
\label{f:B0ALens2D}
\end{figure*}
% *-*-*-*-*-*-*-*-*-*-*-*-*-*-*-*

We see some interesting results for the $A_{\rm Lens}$ in the right hand side of figure  \ref{f:1Df(R)ALB0AL}.  Most notable of these results is that for PLC and PLC+WiggleZ$_{0.1}$ data sets, the deviation of $A_{\rm Lens}$ from unity is reduced, especially so in the latter case
\begin{eqnarray}
 A_{\rm Lens} &=& 1.17\pm0.14,\; (68\%,\, \mbox{Model II:}\;{\rm PLC}),\label{eq:f(R)ALPLC}\\
 A_{\rm Lens} &=& 1.08\pm0.14,\;(68\%,\, \mbox{Model II:}\;{\rm PLC}+{\rm WiggleZ}_{0.1}).\label{eq:f(R)ALWIGZk0p1}
\end{eqnarray}
For the PLC+WiggleZ$_{0.2}$ data set, however, the $2\sigma$ deviation from unity persists 
\begin{equation}
 A_{\rm Lens} = 1.20\pm0.10,\;(68\%,\, \mbox{Model II:}\;{\rm PLC}+{\rm WiggleZ}_{0.2}).\label{eq:f(R)ALWIGZk0p2}
\end{equation}

As shown in figure \ref{f:B0ALens2D}, this is due to a parameter degeneracy between  $A_{\rm Lens}$ and $B_0$ at larger values of $B_0$.  For the PLC and PLC+WiggleZ$_{0.1}$ data set combinations, the mean values for $B_0$ are large, and the resulting value favoured by $A_{\rm Lens}$ is pushed back toward unity. However, when $B_0$ is small, as is the case when using the PLC + WiggleZ$_{0.2}$ data set, a larger value of $A_{\rm Lens}$ is still needed.

%% ------------------------------------------------------------------ %%
\subsection{Model III - $f(R)$ gravity + $A_{\rm Lens}$ + $\sum m_\nu$}

% *-*-*-*-*-*-*-*-*-*-*-*-*-*-*-*
\begin{figure*}[htmb!]
\centering
\includegraphics[scale=0.4]{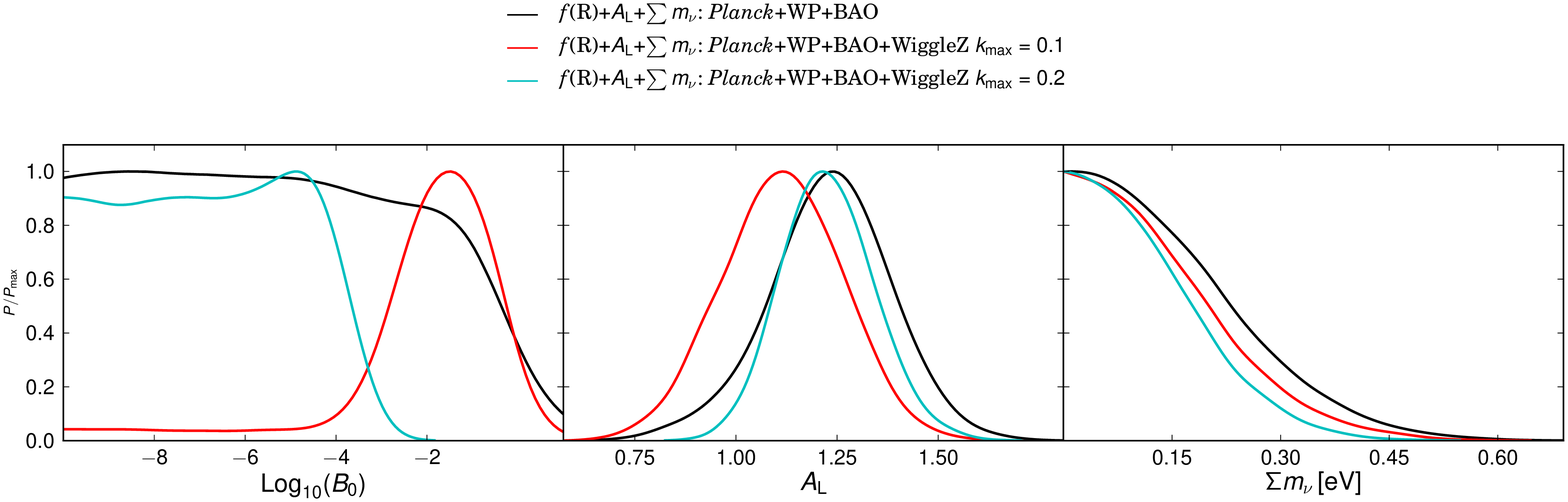}
\caption{\label{f:1DfoR+AL+mnu} 1D posterior distributions from Model III for our three different combinations of data sets. {\it Left panel:} Distributions for ${\rm log}_{10}(B_0)$. The non-zero detection of $B_0$ still persists for the ${\rm PLC}+{\rm WiggleZ_{0.1}}$ data set, as it did in the previous two models.  The PLC+WiggleZ$_{0.2}$ data set still provides a very stringent constraint on $B_0$. {\it Middle panel:}
1D posterior distributions for $A_{\rm Lens}$.  Again, the PLC + WiggleZ$_{0.1}$ (red curve) shows a preference for $A_{\rm Lens}$ very close to its expected value of one, which is not the case for the other two data sets. {\it Right panel:} Distributions for $\sum m_\nu$. The neutrino mass constraint tightens from the PLC constraint as more WiggleZ data is added.}

\end{figure*}
% *-*-*-*-*-*-*-*-*-*-*-*-*-*-*-*

% *-*-*-*-*-*-*-*-*-*-*-*-*-*-*-*
\begin{figure}
\centering
\begin{tabular}{c}
{\includegraphics[width=8cm]{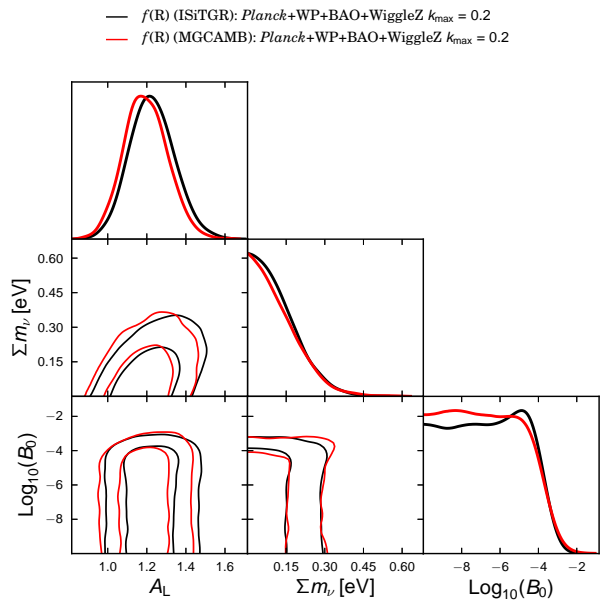}} \\ 
\end{tabular}
\caption{\label{f:ISiTGR_VS_MGCAMB}
Plot showing the consistency of the resulting posterior probability distributions from ISiTGR and MGCAMB.
Moreover, there exists a significant positive correlation between lensing amplitude $A_{\rm Lens}$ and summed neutrino mass $\sum m_{\nu}$.}
\end{figure}
% *-*-*-*-*-*-*-*-*-*-*-*-*-*-*-*

Finally we consider constraints on the cosmological parameters under $f(R)$ gravity where both the 
CMB lensing parameter $A_{\rm Lens}$ and the summed neutrino mass $\sum m_\nu$ are also allowed to vary. Normally constraints on the neutrino mass are computed under the assumption of a $\Lambda$CDM model, but it is important to see how much these constraints degrade when the theory of gravity changes, and also how neutrinos degrade the MG gravity signal. Massive neutrinos affect the late-time matter power spectrum through the suppression of structure, acting similar to warm dark matter (for a recent review, see \cite{RiemerSorensen:2013b}). However, $f(R)$ gravity acts to enhance structure formation (on scales smaller than the Compton wavelength), acting in reverse to the massive neutrino signal. There have been recent results from short baseline experiments suggesting the existence of one or more sterile neutrinos with a mass $\sim1~{\rm eV}$ \cite{Kopp:2011,Mention:2011,Huber:2011,Giunti:2011}, much larger than the current cosmological limit of $0.18~{\rm eV}$ \cite{Riemer-Sorensen:2013}.  Therefore itmay be possible to reconcile these larger neutrino mass detections coming from particle physics experiments with the tight constraints from cosmological data by changing to a MG model. There has already been some research done in this area \cite{Baldi:2013, He:2013,Motohashi:2013}.

We plot the one- and two-dimensional posterior distributions of ${\rm log}_{10}(B_0)$, $A_{\rm Lens}$ and $\sum m_{\nu}$ in figure \ref{f:1DfoR+AL+mnu} and figure \ref{f:ISiTGR_VS_MGCAMB}. As previously stated, we use two independent Einstein-Boltzmann solvers for this model, namely \texttt{ISiTGR} and \texttt{MGCAMB}, and get very consistent results for all the models. The comparison of the results from the two is shown in figure \ref{f:ISiTGR_VS_MGCAMB}. 

The one-dimensional probability distributions for $B_0$ have essentially the same form as in the previous two models, with some small change in the upper limit
\begin{eqnarray}
 {\rm log}_{10}(B_0) &<& -0.75, (95\%, \, \mbox{ Model III:}\;{\rm PLC}),\label{eq:f(R)ALmnuPLC}\\
 {\rm log}_{10}(B_0) &<& -0.24, (95\%, \, \mbox{ Model III:}\;{\rm PLC}+{\rm WiggleZ}_{0.1}),\label{eq:f(R)ALmnuWIGZk0p1}\\
 {\rm log}_{10}(B_0) &<& -3.90, (95\%, \, \mbox{ Model III:}\;{\rm PLC}+{\rm WiggleZ}_{0.2}).\label{eq:f(R)ALmnuWIGZk0p2}
\end{eqnarray}
This change comes about due to the change in the constraint on $\Omega_m$, as the addition of massive neutrinos allows for larger values of $\Omega_m$, which is in opposition to changing from $\Lambda$CDM to $f(R)$ gravity, allowing for smaller values of $\Omega_m$.  We see this effect also on the constraint on the sum of the neutrino mass, which becomes less tight compared with 
the results ($\sum m_{\nu}<0.18$ at $95\%$CL) obtained for the $\Lambda$CDM model \cite{Riemer-Sorensen:2013}
\begin{eqnarray}
 \sum m_{\nu}&<&0.36, (95\%, \,  \mbox{ Model III:}\;{\rm PLC}),\label{eq:f(R)ALmnuPLCmnu}\\
 \sum m_{\nu}&<&0.33, (95\%, \, \mbox{ Model III:}\;{\rm PLC}+{\rm WiggleZ}_{0.1}),\label{eq:f(R)ALmnuWIGZk0p1mnu}\\
 \sum m_{\nu}&<&0.28, (95\%, \, \mbox{ Model III:}\;{\rm PLC}+{\rm WiggleZ}_{0.2}).\label{eq:f(R)ALmnuWIGZk0p2mnu}
\end{eqnarray}

Because of the degeneracy between these three parameters ($\Omega_m$, $B_0$ and $\sum m_\nu$), then the constraint on all of them is degraded. However, even though the upper limit on the neutrino mass is increased, it doesn't bring it up to the level of $\sim1~{\rm eV}$ required by short baseline experiments \cite{Kopp:2011,Mention:2011,Huber:2011,Giunti:2011}.

Secondly, in figure \ref{f:ISiTGR_VS_MGCAMB} we see a small positive correlation between the lensing amplitude $A_{\rm Lens}$ and the sum of the neutrino masses $\sum m_{\nu}$. In Model II model we set $\sum m_{\nu}=0.06~{\rm eV}$ and saw the deviation  of $A_{\rm Lens}$ from unity reduced somewhat.  The positive correlation between $A_{\rm Lens}$ and $\sum m_{\nu}$ leads to an increase in the mean value of $A_{\rm Lens}$ when both are varied, but not by a statistically significant amount,
\begin{eqnarray}
 A_{\rm Lens} &=&\begin{cases}
 1.17\pm0.14,\; (68\%,\, \mbox{Model II:}\;{\rm PLC})\\ 
 1.23\pm0.15,\; (68\%, \, \mbox{ Model III:}\;{\rm PLC}),\label{eq:f(R)ALmnuPLCAL}
 \end{cases}\\
 A_{\rm Lens}&=&\begin{cases}
 1.08\pm0.14,\;(68\%,\,\mbox{Model II:}\;{\rm PLC}+{\rm WiggleZ}_{0.1})\\
 1.11\pm0.15,\;(68\%,\,\mbox{Model III:}\;{\rm PLC}+{\rm WiggleZ}_{0.1}), \label{eq:f(R)ALmnuWIGZk0p1AL}
 \end{cases}\\
 A_{\rm Lens}&=&\begin{cases}
 1.20\pm0.10,\;(68\%,\, \mbox{Model II:}\;{\rm PLC}+{\rm WiggleZ}_{0.2})\\
 1.23\pm0.11,\;(68\%,\, \mbox{Model III:}\;{\rm PLC}+{\rm WiggleZ}_{0.2}).\label{eq:f(R)ALmnuWIGZk0p2AL}
 \end{cases}
\end{eqnarray}

The means and standard deviations of the parameters in this model are given in table \ref{Tab:foR+AL+mnu} in appendix \ref{App:paramconstraints}.

%% ================================================================ %%
\section{Conclusions}
\label{sec:conclusions}
In this paper we studied the parameter constraints in $f(R)$ gravity with Planck CMB data and 
galaxy power spectra from the WiggleZ Dark Energy Survey. We use linear theory predictions for the matter power spectra, and find that the combined data sets give a very tight constraint on the $f(R)$ gravity Compton wavelength parameter, ${\rm log}_{10}(B_0)<-4.07$ at $95\%$ C.L. for the simplest seven parameter model. This represents an order of magnitude improvement over previous analyses. This comes from the ability of large scale structure data to constrain the history of structure formation on different scales. The stringency of our constraints is helped by the assumption non-informative logarithmic prior on $B_0$. We argue that this is the correct approach when testing $f(R)$ gravity models, since we have no prior knowledge for the scale of the Compton wavelength.

Also, we explored the effect of changing the maximum wavenumber of the WiggleZ data which is used the analysis. Interestingly we find that with $k_{\rm max}=0.1h/{\rm Mpc}$ a non-zero Compton wavelength is detected with over $2\sigma$ confidence level. However, as discussed in the previous paragraph, this detection disappears and the parameter is tightly constrained when we increase the wavenumber cutoff to $k_{\rm max}=0.2h/{\rm Mpc}$. This is due to the fact that for $k_{\rm max} = 0.1h/{\rm Mpc}$  the power spectra data has a slight preference for a lower value of the matter density, $\Omega_m$, and hence a higher value of both $\sigma_8$ and $B_0$ are needed fit the CMB data. This difference in best fit between WiggleZ$_{0.1}$ and WiggleZ$_{0.2}$ manifests due to the scale-dependent growth in $f(R)$. Combining a boost in small scale power in the $f(R)$ models with our free linear bias parameter results in a better fit at larger scales by predicting a slightly lower power for the small k-values. However, when more smaller scale data is included, as is the case when $k_{\rm max}=0.2 h/{\rm Mpc}$, the scale-dependent growth of $f(R)$ provides a worse overall fit, and the data disfavours larger values of the Compton wavelength.

In addition, once the CMB lensing amplitude $A_{\rm Lens}$  is included in addition  to the $f(R)$ gravity parameter, $B_0$, a significant correlation between $A_{\rm Lens}$ and $B_0$ is present at larger values of $B_0$, where the growth of structure is more significantly impacted by the MG model. However, it disappears when $B_0$ gets constrained to smaller values by the matter power spectrum data.  
Hence, we can only reduce the deviation of $A_{\rm Lens}$ from the theoretical prediction (unity) with a large modification of gravity, i.e. $B_0>0.1$. 

Finally, we investigate the effect of varying the sum of the neutrino mass $\sum m_\nu$, in an attempt to determine if $f(R)$ can mitigate the tension between cosmological constraints and the large mass detections (1 eV) from small-scale baseline experiments. We vary $A_{\rm Lens}$ and $\sum m_{\nu}$ simultaneously, and find a small positive correlation between the two parameters, $A_{\rm Lens}-\sum m_{\nu}$. A larger neutrino mass makes the deviation of lensing amplitude from unity slightly larger but not at any statistically significant level. We find that the 95\% C.L. on the sum of the neutrino mass increases, due the addition of extra parameters, but not enough to reconcile with results from particle physics experiments. 

%% ================================================================ %%
\acknowledgments
We thank Tamara Davis for useful comments.  JD acknowledges support through a University of Queensland Foundation Research Excellence Award (Tamara Davis, 2012) and that part of the calculations for this work have been performed on the Cosmology Computer Cluster at The University of Texas at Dallas which is funded by the Hoblitzelle Foundation through Mustapha Ishak.  BH is supported by the Dutch Foundation for Fundamental Research on Matter (FOM).

\providecommand{\href}[2]{#2}\begingroup\raggedright\endgroup

\appendix
\section{Marginalised parameter constraints}
\label{App:paramconstraints}

% *-*-*-*-*-*-*-*-*-*-*-*-*-*-*-*
\setlength\tabcolsep{1pt}
\begin{table*}[htb!]
\footnotesize
\centering
\begin{tabular}{|l|c|c|c|}
\hline
&  Model I: & Model I: & Model I:\\ 
& {\bf PLC} & {\bf PLC + WiggleZ$_{0.1}$}  & {\bf PLC + WiggleZ$_{0.2}$} \\
%& & {\bf + } & {\bf + WiggleZ($k_{\rm max}=0.2$)}\\
\hline
Parameters & mean $\pm$ $68\%$ C.L. & mean $\pm$ $68\%$ C.L. & mean $\pm$ $68\%$ C.L.\\ \hline
$100\Omega_b h^2$ & 2.238$\pm$0.028 & 2.243$\pm$0.027 & 2.217$\pm$0.025 \\
$\Omega_c h^2$ & 0.1172$\pm$0.0017 & 0.1171$\pm$0.0016 & 0.1184$\pm$0.0016 \\
$100\theta$ & 1.04174$\pm$0.00057 & 1.04178$\pm$0.00057 & 1.04156$\pm$0.00056 \\
$\tau$ & 0.091$\pm$0.013 & 0.089$\pm$0.013 & 0.092$\pm$0.013 \\
$n_s$ & 0.9675$\pm$0.0057 & 0.9679$\pm$0.0056 & 0.9642$\pm$0.0056 \\
$\log(10^{10} A_s)$ & 3.086$\pm$0.026 & 3.083$\pm$0.025 & 3.092$\pm$0.025 \\
$\log_{10}(B_0)$ & $<-0.44$~(95$\%$CL) & $-1.42^{+1.55}_{-1.66}$~(95$\%$CL) & $<-4.07$~(95$\%$CL) \\
\hline
$\sigma_8$ & 1.028$\pm$0.136 & 1.101$\pm$0.083 & 0.836$\pm$0.018 \\
$\Omega_{m}$ & 0.297$\pm$0.010 & 0.296$\pm$0.010 & 0.305$\pm$0.009 \\
$H_0 [\mathrm{km}/\mathrm{s}/\mathrm{Mpc}]$ & 68.72$\pm$0.79 & 68.80$\pm$0.77 & 68.04$\pm$0.72 \\
\hline
\hline
$\chi^2_{\rm min}/2$ & 4903.084 & 4998.130 & 5131.392 \\
\hline
\end{tabular}
\caption{
Mean values and $68\%$ confidence limits for standard primary/derived $\Lambda$CDM parameters and $95\%$ 
confidence limits for ${\rm log}_{10}(B_0)$ in $f(R)$ gravity.  We should note that there are 280 additional data points when using the PLC + WiggleZ$_{0.2}$ data set compared to the PLC + WiggleZ$_{0.1}$ data set.}
\label{Tab:foR}
\medskip
\begin{tabular}{|l|c|c|c|}
\hline
&  Model II: & Model II: & Model II:\\ 
& {\bf PLC} & {\bf PLC + WiggleZ$_{0.1}$}  & {\bf PLC + WiggleZ$_{0.2}$} \\
\hline
Parameters & mean $\pm$ $68\%$ C.L. & mean $\pm$ $68\%$ C.L. & mean $\pm$ $68\%$ C.L.\\ \hline
$100\Omega_b h^2$ & 2.250$\pm$0.027 & 2.248$\pm$0.027 & 2.242$\pm$0.027 \\
$\Omega_c h^2$ & 0.1166$\pm$0.0017 & 0.1170$\pm$0.0017 & 0.1175$\pm$0.0016 \\
$100\theta$ & 1.04186$\pm$0.00058 & 1.04182$\pm$0.00057 & 1.04176$\pm$0.00057 \\
$\tau$ & 0.087$\pm$0.013 & 0.088$\pm$0.013 & 0.086$\pm$0.013 \\
$n_s$ & 0.9697$\pm$0.0059 & 0.9687$\pm$0.0058 & 0.9678$\pm$0.0058 \\
$\log(10^{10} A_s)$ & 3.078$\pm$0.025 & 3.080$\pm$0.025 & 3.078$\pm$0.025 \\
$\log_{10}(B_0)$ & $\cdots$~$\cdots$ & $-2.05^{+3.03}_{-6.56}$~($95\%$CL) & $<-4.02$~($95\%$CL) \\
$A_L$ & 1.17$\pm$0.14 & 1.08$\pm$0.14 & 1.20$\pm$0.10 \\
\hline
$\sigma_8$ & 0.904$\pm$0.129 & 1.053$\pm$0.122 & 0.827$\pm$0.019 \\
$\Omega_{m}$ & 0.293$\pm$0.010 & 0.295$\pm$0.010 & 0.298$\pm$0.010 \\
$H_0 [\mathrm{km}/\mathrm{s}/\mathrm{Mpc}]$ & 69.07$\pm$0.81 & 68.91$\pm$0.79 & 68.65$\pm$0.77 \\
\hline
\hline
$\chi^2_{\rm min}/2$ & 4902.900 & 4998.066 & 5129.378 \\
\hline
\end{tabular}
\caption{Mean values and $68\%$ ($95\%$ for ${\rm log}_{10}(B_0)$) confidence limits for parameters in $f(R)+A_{\rm Lens}$ model.
The dots in the first column indicate no upper bound is found for ${\rm log}_{10}(B_0)$ in the parameter space we sampled.}
\label{Tab:foR+AL}
\medskip
\begin{tabular}{|l|c|c|c|}
\hline
&  Model III: & Model III: & Model III:\\ 
& {\bf PLC} & {\bf PLC + WiggleZ$_{0.1}$}  & {\bf PLC + WiggleZ$_{0.2}$} \\
\hline
Parameters & mean $\pm$ $68\%$ C.L. & mean $\pm$ $68\%$ C.L. & mean $\pm$ $68\%$ C.L.\\ \hline
$100\Omega_b h^2$ & 2.259$\pm$0.030 & 2.254$\pm$0.029 & 2.246$\pm$0.028 \\
$\Omega_c h^2$ & 0.1155$\pm$0.0021 & 0.1162$\pm$0.0020 & 0.1169$\pm$0.0019 \\
$100\theta$ & 1.04199$\pm$0.00059 & 1.04190$\pm$0.00059 & 1.04182$\pm$0.00058 \\
$\tau$ & 0.088$\pm$0.013 & 0.089$\pm$0.013 & 0.087$\pm$0.013 \\
$n_s$ & 0.9721$\pm$0.0066 & 0.9707$\pm$0.0063 & 0.9692$\pm$0.0062 \\
$\log(10^{10} A_s)$ & 3.078$\pm$0.026 & 3.080$\pm$0.026 & 3.078$\pm$0.026 \\
$\log_{10}(B_0)$ & $<-0.75$~($95\%$CL) & $<-0.24$~($95\%$CL) & $<-3.90$~($95\%$CL) \\
$\sum m_{\nu}$ [eV] & $<0.36$~($95\%$CL) & $<0.33$~($95\%$CL) & $<0.28$~($95\%$CL) \\
$A_L$ & 1.23$\pm$0.15 & 1.11$\pm$0.15 & 1.23$\pm$0.11 \\
\hline
$\sigma_8$ & 0.866$\pm$0.123 & 1.027$\pm$0.120 & 0.814$\pm$0.029 \\
$\Omega_{m}$ & 0.296$\pm$0.011 & 0.298$\pm$0.010 & 0.301$\pm$0.011 \\
$H_0 [\mathrm{km}/\mathrm{s}/\mathrm{Mpc}]$ & 68.73$\pm$0.90 & 68.61$\pm$0.88 & 68.40$\pm$0.87 \\
\hline
\hline
$\chi^2_{\rm min}/2$ & 4903.634 & 4997.893 & 5129.571 \\
\hline
\end{tabular}
\caption{Mean values and $68\%$ ($95\%$ for ${\rm log}_{10}(B_0)$ and $\sum m_{\nu}$) confidence 
limits for parameters in $f(R)+A_{\rm Lens}+\sum m_{\nu}$ model.}
\label{Tab:foR+AL+mnu}
\end{table*}
\vfill

\end{document}